\documentclass[aps, prd, preprint, superscriptaddress, amsmath, amssymb, showpacs]{revtex4-1}
\usepackage{dcolumn}
\usepackage{graphicx}
\usepackage{float}
\usepackage{physics}
\usepackage[colorlinks=true, allcolors=blue]{hyperref}
\UseRawInputEncoding
\begin{document}
	\title{Effects of tensor spin polarization on the chiral restoration and deconfinement phase transitions \\
	}
	
	\author{Yan-Ru Bao}
	\affiliation{College of Science, China Three Gorges University, Yichang 443002, China}
	
	\author{Sheng-Qin Feng}
	\email{Corresponding author: fengsq@ctgu.edu.cn}
	\affiliation{College of Science, China Three Gorges University, Yichang 443002, China}
	\affiliation{Center for Astronomy and Space Sciences and Institute of Modern Physics, China Three Gorges University, Yichang 443002, China}
	\affiliation{Key Laboratory of Quark and Lepton Physics (MOE) and Institute of Particle Physics, \\
		Central China Normal University, Wuhan 430079, China}
	
	\date{\today}

	\begin{abstract}
		Abstract:  Effects of tensor spin polarization (TSP) on the chiral restoration and deconfinement phase transitions are studied in Polyakov loop extended Nambu-Jona-Lasinio (PNJL) model. For chiral phase transition, the higher the polarized degree of quark-antiquark pairs under the strong magnetic field, the higher the phase transition temperature. The TSP corrects the position of the critical end point. The small impact of TSP on the phase transition temperature is found for the deconfinement phase transition. On the other hand, we divide the phase space into three ranges based on the phase diagram obtained from the PNJL model: the confinement phase with chiral symmetry broken, the deconfinement phase with restored chiral symmetry, and the confinement phase with restored chiral symmetry (quarkyonic phase). It is found that TSP has only a very small effect on the anisotropic pressure in the deconfined phase with chiral symmetry restored and the quarkyonic phase, but it has a very strong effect on the anisotropic pressure in the confined phase with chiral symmetry broken. This is because TSP is closely related to chiral symmetry. The restoration of chiral symmetry means the dissociation of spin polarization condensate.
	\end{abstract}
	
	\maketitle
	
	\section{Introduction}\label{sec:01_intro}
	
	The properties of QCD mater under strong magnetic field background have attracted widespread interest among researchers \cite{RN1, RN2, RN3, RN4, RN5, RN6, RN7, RN8}. In the early Universe, the magnetic field may reach the order of $10^{22}~~\mathrm{G}$ \cite{RN9, RN10}. The surface magnetic field of magnetars can reach $10^{14}$-$10^{15}~~\mathrm{G}$, while its internal magnetic field can reach $10^{18}$-$10^{20} ~~\mathrm{G}$ \cite{RN11, RN12}. And the extremely high magnetic fields can be generated in noncentral heavy ion collision experiments. The magnetic field created in Relativistic Heavy Ion Collider experiments can reach $\sqrt{eB} \sim$ 0.1 GeV, while in the LHC, the magnetic field intensity can reach $\sqrt{eB} \sim$ 0.5 GeV. Although the magnetic field is an external field with a short lifetime (on the order of 1 fm/$\textit{c}$), and the existence of the quark-gluon plasma (QGP) medium response effect is significant, it greatly delays the decay of these time-dependent magnetic fields \cite{RN13, RN14, RN15, RN16}. Therefore, it seems reasonable to assume the existence of a constant external magnetic field in many cases. The addition of strong magnetic field will make the phase diagram and phase structure of QCD more complex and interesting, leading to the emergence of many new phenomena, such as chiral magnetic effect \cite{RN17, RN18, RN19, RN20}, magnetic catalysis in vacuum (MC) \cite{RN21, RN22, RN23}, and inverse magnetic catalysis \cite{RN24, RN25, RN26, RN27, RN28, RN29} around chiral phase transitions.
	
	The magnetic field also induces spin polarization, which is the condensate of quark-antiquark pairs with parallel spins. As shown in Ref. \cite{RN30}, the tensor-type interaction $ ( \bar { \psi } \Sigma ^ { 3 } \psi ) ^ { 2 } + ( \bar { \psi } i \gamma ^ { 5 } \Sigma ^ { 3 } \psi ) ^ { 2 }$ induces spin polarization $ < \bar { \psi } i \gamma ^ { 1 } \gamma ^ { 2 } \psi >$ which is analogous to the form of an anomalous magnetic moment \cite{RN30, RN31} of quarks developed in the presence of a magnetic field. Note that the tensor polarization operator $ < \bar { \psi } \sigma ^ { 1 2 } \psi >$ is also known as the tensor spin polarization (TSP) operator or spin density, because $ < \bar { \psi } \sigma ^ { 1 2 } \psi > = < \psi ^ { \dagger } \gamma ^ { 0 } \Sigma^3 \psi >$ and with ${\Sigma}^3=\Bigl(\begin{smallmatrix}
		{\sigma}^3 & 0 \\
		0 & {\sigma}^3
	\end{smallmatrix}\Bigr)$
	, $ \sigma ^ { 3 } = - i \sigma ^ { 1 } \sigma ^ { 2 }$. By projecting the quark spinors $ \psi$ into the spin subspace $ \psi = \psi _ { \uparrow } + \psi _ { \downarrow }$, one can obtain $ \bar { \psi } \sigma ^ { 1 2 } \psi \sim \langle \bar { \psi } _ { \uparrow } \psi _ { \uparrow } \rangle - \langle \bar { \psi } _ { \downarrow } \psi _ { \downarrow } \rangle$, which measures the difference between spin-up quark pairing and spin-down quark pairing \cite{RN7, RN32}.
	
The effects of TSP on the dynamic generation of quark magnetic moments in nonequilibrium quark matter, the magnetic properties of QCD matter, pion mass, and the chiral phase transition in the (2+1)-flavor Nambu-Jona-Lasinio (NJL) model have been investigated \cite{RN7, RN31, RN32}. We know that the NJL model can only discuss issues such as chiral symmetry breaking and restoration phase transition, while the Polyakov loop extended Nambu-Jona-Lasinio (PNJL) model can analyze both chiral and deconfinement phase transition simultaneously.
	
	In recent years, the anisotropy induced by magnetic fields has also been widely studied \cite{RN33, RN34}. The destruction of rotational symmetry by magnetic fields leads to the anisotropy of energy-momentum tensors (EMTs). If the spatial elements of EMTs are interpreted as the pressure generated by the response of the thermodynamic potential of the system to compression in the corresponding direction, there exists a difference caused by the orientation of the magnetic field in the local rest framework \cite{RN35}. It has been proven that the derivative of the partition function obtained under constant magnetic flux corresponds to spatial elements where the directional difference of EMTs becomes apparent. Usually, these different elements are referred to as longitudinal ($ P _ { \parallel }$) and transverse ($ P _ { \bot }$) pressures \cite{RN26}. These quantities will affect the equation of state of strongly interacting substances.
	
	Once the anisotropy of pressure under magnetic field background is considered, many studies on the equation of state of dense stars will yield new results \cite{RN35, RN36, RN37, RN38, RN39, RN40, RN41, RN42, RN43, RN44, RN45, RN46, RN47}. Because of the fact that both TSP and anisotropic pressure are caused by magnetic fields, it is of great research significance to focus on the influence of TSP on anisotropic pressure. In addition, under the magnetic field background, the original isotropic fermion vertices split into longitudinal and transverse fermions in the new tensor channel \cite{RN28}; it is expected that this anisotropy will be reflected in pressure, and TSP will promote the anisotropy of pressure.
	
	The impacts of TSP on the chiral restoration phase transition and deconfinement phase transition in the PNJL model are investigated in the paper. Then, the impacts of TSP on the anisotropic pressure under three different phase ranges are studied. This paper is organized as follows. The two-flavor PNJL model with tensor channel is introduced in Sec. II. In Sec. III, we will study the distributions of TSP under different backgrounds, the influences of TSP on chiral restoration and deconfinement phase transition, and the effect of TSP on the anisotropy of pressure under three different phases. Finally, we make the summaries and conclusions in Sec. IV.

	\section{The Two-flavors PNJL model with TSP}\label{sec:02 setup}
	
	The destruction of rotational symmetry by uniform magnetic field leads to the separation of longitudinal and transverse-fermion modes along the direction of the magnetic field \cite{RN28, RN30}. This separation leads to an effective splitting of the coupling in the gluon exchange interaction on which the NJL model is usually based. Therefore, this splitting can be reflected in the four-fermion coupling of the QCD effective field NJL model. By using Fierz identities \cite{RN30, RN32, RN48} in a magnetic field, we can obtain the Lagrangian of the scalar and tensor interactions of the two-flavor PNJL model as
	\begin{equation}\label{eq:001}
		\begin{aligned}
			\mathcal{L} = & \bar{\psi} (i\gamma _{\mu}D^{\mu}-m_0+\gamma ^0\mu_q)\psi+G_s[(\bar{\psi}\psi)^2 +(\bar{\psi}i\gamma ^5\psi)^2] \\	& +G_t[(\bar{\psi}\Sigma^3\psi)^2+(\bar{\psi}i\gamma^5\Sigma^3\psi)^2]-\mathcal{U}(\Phi, \bar{\Phi} ) ,
		\end{aligned}
	\end{equation}
	where $\psi=(u, d)^T$ is two-flavor quark field with $\hat{m} =\textrm{diag}(m_u, m_d)$. Because of the spin symmetry of light quarks, the current quark mass is $m_0=m_u=m_d$. $\Sigma^3=\frac{i}{2} [\gamma^1, \gamma^2]=i\gamma^1\gamma^2$ is the spin operator.  In addition, the covariant derivative $D^{\mu}=\partial^{\mu}+i\hat{q_{f}}A^{\mu}-i\mathcal{A}^{\mu}$ couples quarks to the two fields: (1) the magnetic field $\mathbf{B} = \nabla\times \mathbf{A}$ and (2) the temporal gluon field $\mathcal{A}^{\mu}=\delta^{\mu}_{0}\mathcal{A}^0$ with $\mathcal{A}^{0}=g\mathcal{A}^{0}_{a}\lambda_{a}/2=-i\mathcal{A}_{4}$. The gauge coupling $g$ is linked with the SU(3) gauge field $\mathcal{A}^{0}_{a}$ to define $\mathcal{A}^{\mu}(x)$, $\hat{q_{f}}=\textrm{diag}(q_u, q_d)=\textrm{diag}(\frac{2}{3}e, -\frac{1}{3}e )$ is the quark charge matrix in flavor space, and $\lambda_a$ are the Gell-Mann matrices in color space. In the external electromagnetic field $A^{\mu}=(0, 0, Bx_1, 0)$, a constant and homogenous magnetic field of magnitude $B$ points toward the $x_3$ direction.
	
	The second term in Eq. (1) is the traditional scalar channel, which produces a dynamical quark mass. The third term of Eq. (1) is the tensor channel, which preserves chiral symmetry and rotational symmetry along the direction of the magnetic field. The tensor channel is closely related to spin interactions and induces spin polarization. In the magnetic field background, the running coupling constant is divided into longitudinal ($ g _ { \parallel }$) and transverse ($ g _ { \bot }$) components \cite{RN30}. The coupling coefficients $G_s$ and $G_t$ of the NJL interaction related to quark-gluon vertex coupling can be determined by $G_s=(g_{\parallel}^2+g_{\bot}^2)/\Lambda^2$ and $G_t=(g_{\parallel}^2-g_{\bot}^2)/\Lambda^2$. By setting $G_t/G_s=\alpha$, one can obtain $g_{\parallel}=g_{\bot}(\alpha=0)$ at zero magnetic field, and $g_{\parallel}\gg g_{\bot}(\alpha\to 1)$ as $eB\to \infty$. In the following, we will choose the cases of $\alpha=0$, $\alpha = \frac{1}{2}$ and $\alpha = 1$.
	
	By using the mean-field approximation, one can obtain the Lagrangian density as
	\begin{equation}\label{eq:002}
		\mathcal{L} _{MF}=\bar{\psi}(i\gamma^{\mu}D_{\mu}-M+\gamma^0\mu_q-i\xi\gamma^{1}\gamma^{2})\psi-\frac{\sigma^2}{4G_s}-\frac{\xi^2}{4G_t}-\mathcal{U}(\Phi, \bar{\Phi}),
	\end{equation}
	where $M=m_0+\sigma$ is the dynamical quark mass, and $\sigma=-2G_s\left \langle\bar{\psi}\psi\right \rangle $ is the chiral condensate. As mentioned earlier, tensor channels are closely related to spin-spin interactions. Under the background of a magnetic field, quark-antiquark pairs with opposite spin and opposite charge undergo orderly arrangement, resulting in tensor spin polarization condensate
	\begin{equation}\label{eq:003}
		\xi=-2G_t\left \langle\bar{\psi}\Sigma^3\psi\right \rangle.
	\end{equation}
	
	The fourth term of Eq. (1) is the Polyakov potential $\mathcal{U}(\Phi, \bar{\Phi})$ associated with the deconfinement phase transition \cite{RN49}, where $\Phi$ is the order parameter describing the deconfinement phase transition. When $\Phi\to 0$, the system is considered to be in the confinement phase, while when $\Phi\to 1$, the system is considered to be in the deconfinement phase. The Polyakov potential $\mathcal{U}(\Phi, \bar{\Phi})$ is given as
	\begin{equation}\label{eq:004}
		\frac{\mathcal{U}(\Phi, \bar{\Phi} )}{T^4} =-\frac{1}{2} A(T)\bar{\Phi}\Phi+B(T)\ln\left \{ {1-6\bar{\Phi}\Phi+4(\bar{\Phi}^3+\Phi^3)-3(\bar{\Phi}\Phi)^2}\right \},
	\end{equation}
	where the Polyakov potential $\mathcal{U}(\Phi, \bar{\Phi})$ is related to the Z(3) center symmetry. By simulating the deconfinement at finite temperature, one can obtain \cite{RN50} the coefficients as
	\begin{equation}
		A(T)=a_0+a_1(\frac{T_0}{T} )+a_2(\frac{T_0}{T} )^2,
	\end{equation}
	\begin{equation}
		B(T)=b_3(\frac{T_0}{T} )^3.
	\end{equation}\
	The different parameters \cite{RN50} of Eqs. (5) and (6) are given in Table I.
	\begin{table}[h]
	\caption{Parameters set for Polyakov potential.}
	\begingroup
	\renewcommand*{\arraystretch}{1.5}
	\setlength{\tabcolsep}{8pt} 
	\begin{tabular} {cccccc}
		\hline \hline
		$ a_0 $ & $ a_1 $ &  $ a_2 $   & $ b_3 $ & $ T_0 $ (MeV)  \\
		\hline
		$ 3.51 $ & $ -2.47$ & $ 15.2 $ & $ -1.75 $ & $ 270 $  \\
		\hline\hline
	\end{tabular}
	\endgroup
	\label{Table_parameters}
\end{table}

The effective potential at finite temperature and chemical potential obtained by the standardized process is
\begin{equation}\label{eq:007}
	\begin{aligned}
		\Omega=&\frac{\sigma ^2}{4G_s} +\frac{\xi ^2}{4G_t} +\mathcal{U}(\Phi, \bar{\Phi} )	-3\sum_{n, f, s}^{}\frac{\left | q_fB \right | }{2\pi} \int_{-\infty }^{\infty}\frac{dp_z}{2\pi} \epsilon _{n, f, s} \\  &
		-T\sum_{n, f, s}^{}\frac{\left | q_fB \right | }{2\pi } \int_{-\infty }^{\infty }\frac{dp_z}{2\pi }[T\ln(1+g^-)+T\ln(1+g^+)],
	\end{aligned}
\end{equation}\
where
\begin{equation}\label{eq:008}
	g^-(\Phi, \bar{\Phi})=1+3 (\Phi+\bar{\Phi} \exp (\frac{-E _{n, f, s}^{(-)}}{T}))\exp(\frac{-E _{n, f, s}^{(-)}}{T} ) +\exp(\frac{-3E _{n, f, s}^{(-)}}{T} ),
\end{equation}\
\begin{equation}\label{eq:009}
	g^+(\Phi, \bar{\Phi})=1+3( \bar{\Phi}+\Phi \exp  ( \frac{-E _{n, f, s}^{(+)}}{T} )  )\exp
(\frac{-E _{n, f, s}^{(+)}}{T} ) +\exp  (\frac{-3E _{n, f, s}^{(+)}}{T} ),
\end{equation}\
where $E_{n, f, s}^{(\pm)}=\epsilon _{n, f, s}\pm\mu_q$, and the dispersion relation of quarks with TSP \cite{RN31, RN32} is given by
\begin{equation}
	\epsilon _{n, f, s}^2=\begin{cases}p_z^2+(\sqrt{M^2+2n\left | q_f \right |B }-s\xi )^2, n\ge 1, \\p_z^2+(M+\xi )^2, n=0, \end{cases}
\end{equation}
where the summation of $n$ is taken over all Landau levels. $f=u, d$ correspond to flavor quantum number, and the $s=\pm 1$ corresponds to the different spin projections. Note that since the fermion in the lowest Landau level has only one spin projection, no splitting is present in the $n$ = 0 case. But for the excited Landau levels $n\ge 1$, the spectrum of the quasiquarks exhibits a Zeeman splitting ($s=\pm 1$) due to the tensor spin condensation $\xi$. One can obtain the gap equations as
\begin{equation}
	\frac{\partial \Omega}{\partial M} =\frac{\partial \Omega}{\partial\xi}=\frac{\partial \Omega}{\partial \Phi}=\frac{\partial \Omega}{\partial \bar{\Phi} }=0.
\end{equation}

In order to ensure that the thermodynamic potential in vacuum is zero, we define the normalized thermodynamic potential as the effective potential \cite{RN7}
\begin{equation}
	\Omega_{eff}(T, \mu, eB)=\Omega(T, \mu, eB)-\Omega(0, 0, eB).
\end{equation}

The expressions \cite{RN33} of transverse pressure and longitudinal pressure are
\begin{equation}
	P_{\parallel }(T, \mu , eB)=-\Omega_{eff}(T, \mu , eB),
\end{equation}
\begin{equation}
	P_{\bot  }(T, \mu , eB)=P_{\parallel }(T, \mu , eB)-\mathcal{M}eB,
\end{equation}
where the magnetization $\mathcal{M}=-\frac{\partial \Omega_{eff}}{\partial (eB)}$.

It can be found that the energy integral term in the thermodynamic potential equation (7) is ultraviolet divergent. The renormalization scheme cannot be used to eliminate the divergence because of the dotlike interaction between quarks. Therefore, it is necessary to use the appropriate regularization scheme to eliminate UV divergence. We use the Pauli-Villars(PV) regularization scheme with gauge covariance to eliminate the divergence in this paper. The key of the PV regularization scheme is to replace integration with summation after introducing the normalized energy. The normalized energy is given \cite{RN51} as

\begin{equation}
	\epsilon _{n, f, s, i(\textrm{PV})}^2=    \begin{cases}        P_z^2+(\sqrt{M^2+2n\left | q_f \right |B+a_i\Lambda^2 }-s\xi  )^2, n \ge  1, \\        P_z^2+(\sqrt{M^2+a_i\Lambda^2}+\xi )^2, n =  0 ,    \end{cases}
\end{equation}
\begin{equation}
	\sum_{n, f, s}^{} \int_{-\infty }^{\infty } \frac{dp_z}{2\pi } \epsilon _{n, f, s}\to \sum_{n, f, s}^{} \int_{-\infty}^{\infty}\frac{dp_z}{2\pi} \sum_{i=0}^{N} c_i\epsilon _{n, f, s, i(\textrm{PV})} ,
\end{equation}
where the parameters of the PV regularization scheme are given as $\mathit{N}$ = 3, $a_i=\left \{ 0, 1, 2, 3 \right \} $ and $c_i=\left \{ 1, -3, 3, -1 \right \} $, and it  satisfies the formula $\sum_{i=0}^{N} c_{i}\left(M^{2}+a_{i} \Lambda^{2}\right)^{L}=0 $ for $L = 0,1,...,N-1$ \cite{RN29}. By fitting the vacuum values such as the pion-decay constant $f_\pi =93$ MeV and chiral condensation $ \langle \bar{\psi}\psi   \rangle =(-250\ \mathrm{ MeV})^3$, one can obtain \cite{RN51} the relevant parameters such as $G_s = 3.44\ \mathrm{GeV} ^{-2}$, $\Lambda = 1127\ \mathrm{MeV} $, and $m_0 = 5\ \mathrm{MeV}$ at $G_t=0$.

\section{NUMERICAL RESULTS}

\subsection{Tensor spin polarization and chiral condensation}

\begin{figure}[H]
	\centering
	\includegraphics[width=0.55\textwidth]{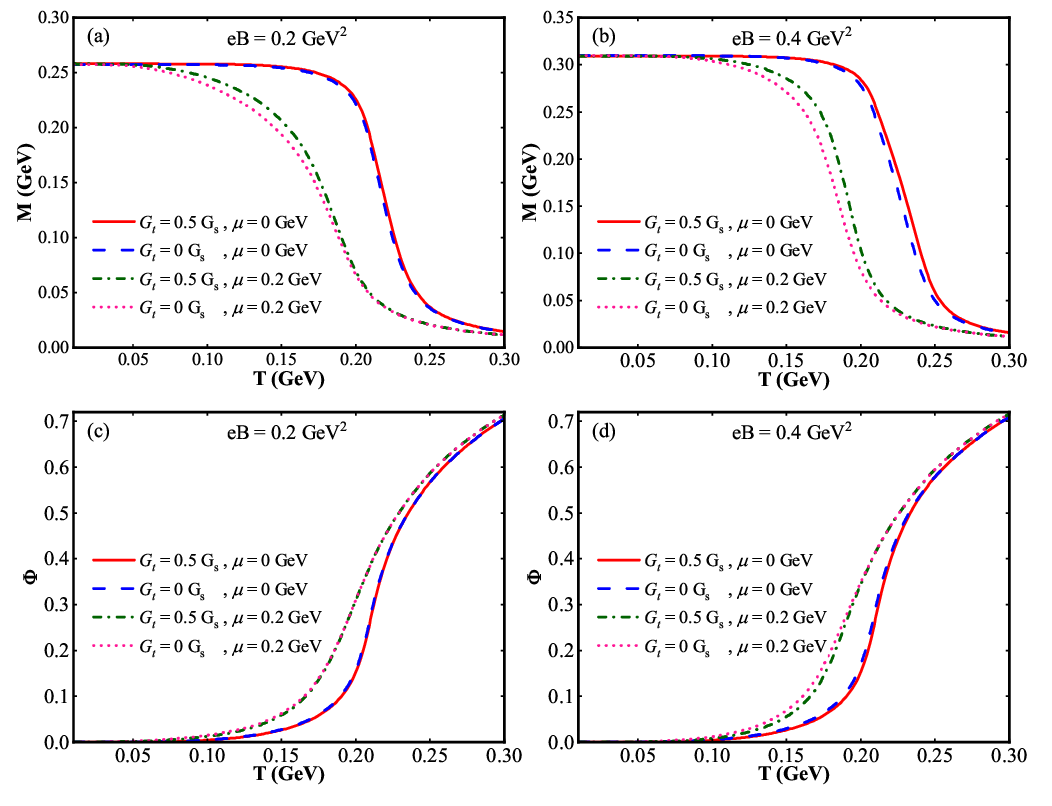}
	\caption{\label{fig1} The dynamical quark mass $M$ and Polyakov loop $\Phi$ as functions of temperature $T$ with different external magnetic field ($eB$ = 0.20  and $eB$ = 0.40 GeV$^{2}$) and different set of spin-spin interaction coupling constants ($G_t$ = 0 $G_s$ and $G_t$ = 0.5 $G_s$ for $\mu$ = 0  and $\mu$ = 0.20 GeV).Figs. 1(a) and 1(b) are for the dynamical quark mass $M$ as functions of temperature $T$ for $eB$ = 0.2  and $eB = 0.4 \textrm{GeV}^{2}$, respectively, while Figs.1(c) and 1(d) are the same as Figs. 1(a) and 1(b), but for Polyakov loop $\Phi$ as functions of temperature $T$.}
\end{figure}

First,  we will investigate the effects of TSP on the chiral and deconfinement phase transitions. The temperature dependence of order parameters $M$ and $\Phi$ under different magnetic fields and chemical potentials are shown in Fig. 1. As mentioned earlier, when $G_t$ is zero, TSP contribution is zero, but when $G_t=0.5\ G_s$ is not zero, TSP contribution is not zero. The temperature dependences of chiral order parameters $M$ in the case of $eB$ = 0.2  and $eB = 0.4 \textrm{GeV}^{2}$ are manifested in Figs. 1(a) and 1(b). The chiral symmetry is broken with $M\ne 0$ at low temperature, and the chiral symmetry is restored with $M\to 0$ at high temperature. It is found that the chiral phase transition temperature increases by considering the influence of TSP. The main reason is that TSP will provide a nonzero magnetic moment for the quasiparticle when the quark obtains the dynamical mass. The magnetic moment produced by the spin polarization under the action of the magnetic field will increase the dynamical mass of the quasiparticle, which induces the MC effect. This MC characteristic is more significant in the high-temperature region.

Comparing Figs. 1(a) and  1(b), we find that the increase of the phase transition temperature is more significant under higher magnetic fields ($eB=0.40 \textrm{GeV} ^2$), which also means that the spin polarization effect is more significant under higher magnetic fields.
The deconfinement phase transitions at $\mu$ and $\mu=0.20 \textrm{GeV} $ are shown in Figs. 1(c) and 1 (d), respectively. It is found that the impact of TSP increases with the magnetic field, but slightly enhances the deconfinement phase transition temperature.

In the $T$-$eB$ plane of Fig. 2, the corresponding temperature range is $0.01 \le T\le 0.25 \textrm{GeV} $, and the magnetic field range is within $0.01 le eB\le 0.40 \textrm{GeV} ^2$. Figs. 2(a) and 2(b) display the contour plots of the spin polarization condensation $\xi=-2G_t\left \langle \bar{\psi }\Sigma^3\psi \right \rangle $ with $\mu=0 $ and $\mu=0.20 \textrm{GeV} $, respectively.
\begin{figure}[H]
	\centering
	\includegraphics[width=0.85\textwidth]{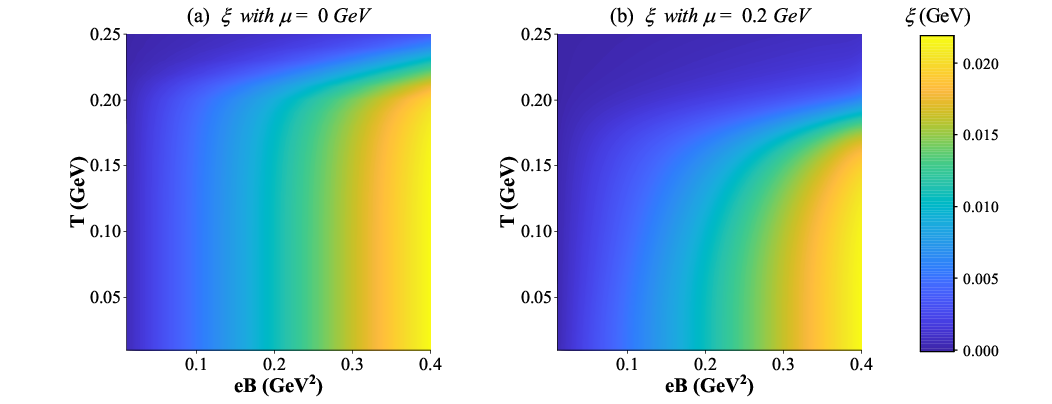}
	\caption{\label{fig2}  contour plots of the distribution of spin polarization condensate $\xi$ for (a) $\mu$ = 0  and (b) $\mu$ = 0.20 GeV in the $T$-$eB$ plane.}
\end{figure}

\begin{figure}[H]
	\centering
	\includegraphics[width=0.85\textwidth]{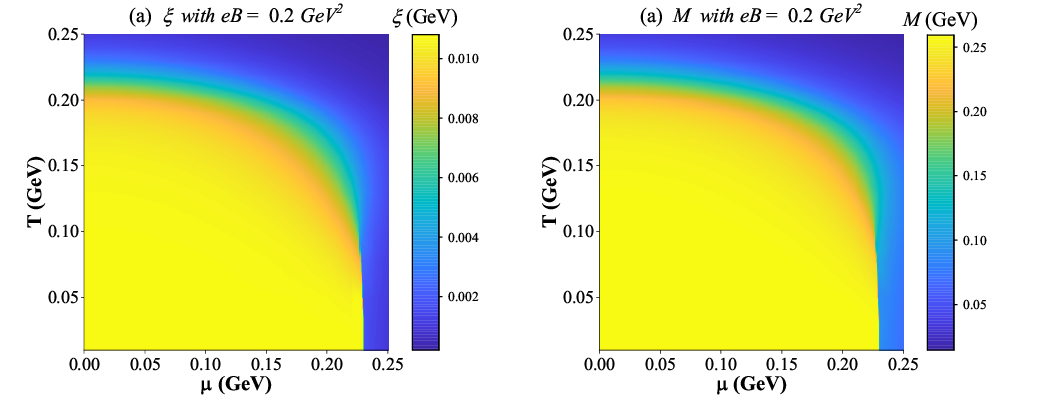}
	\caption{\label{fig3} contour plots of the distribution of (a) spin polarization condensate $\xi$ and (b) dynamical mass $M$ for $eB$ = 0.20 GeV$^{2}$ in the $T$-$\mu$ plane.}
\end{figure}

The spin polarization condensate $\xi$ decreases with the increase of temperature in the $T$-$eB$  plane shown in Figs. 2(a) and 2(b). This shows that the thermal background is not conducive to the formation of quark-antiquark pairs, which leads to the inhibition of TSP production. Figures 2(a) - and 2(b) show that $\xi$ increases with the magnetic field under different chemical potentials. This is because the charged quark-antiquark pairs are easier to be polarized under strong magnetic field.

Figures 3(a) and 3(b) show the distribution plots of spin polarization condensate $\xi$ and dynamical quark mass $M$ with $eB$ = 0.20 GeV$^{2}$ in the $T$-$\mu$ plane. It is worth noting that due to the close relationship between spin polarization condensate $\xi$ (also called dynamical quark moment) and dynamical quark mass $M$ \cite{RN30, RN31}, the $T$-$\mu$ distribution diagrams of $M$ and $\xi$ are very similar. The distribution of $\xi$ shows a continuous change at low chemical potentials and a sharp drop at high chemical potentials, which is consistent with the behavior of the order parameter $M$ during chiral phase transition. Once quarks obtain a dynamical mass, they should also obtain a tensor spin polarization. This effect has also been reported in massless QED and in a one-flavor NJL model \cite{RN30, RN52, RN53,RN54}. From the view of symmetry, once the chiral symmetry is dynamically broken, there is no symmetry protecting the TSP, because a nonvanishing value of the latter breaks exactly the same symmetry.

\subsection{Phase diagram}

The $T$-$\mu$ phase diagram of chiral and deconfinement phase transition with and without TSP ($G_t=0\ G_s$ and $G_t=0.5\ G_s$)  under different magnetic fields are manifested in Fig. 4. It is found that the crossover occurs at high temperature and small chemical potentials $\mu$ while the first-order phase transition happens at low temperatures $T$ and large chemical potential $\mu$. The influences of TSP on the phase diagrams of the deconfinement phase transition and chiral phase transition can be summarized as follows: (1) In general, considering TSP, it has little effect on the deconfinement phase diagram, no matter whether it is a large magnetic field ($eB$ =0.40 GeV$^{2}$) or a small magnetic field ($eB$ =0.20 GeV$^{2}$). (2) TSP has great influence on the chiral phase diagram. In our chiral phase diagram, there are the crossover phase transition region and the first-order phase transition region. Considering the contribution of TSP, the influence on the first-order phase transition is larger and the influence on the crossover phase transition is smaller. However, with the increasing of magnetic field, the influence of TSP on the phase diagram will increase. When the magnetic field rises from 0.2 to $eB$ =0.40 GeV$^{2}$, the line of the first-order phase transition line becomes longer as the magnetic fields become stronger. The results are consistent with those of Ref.\cite{RN55}.

It is worth noting that the confinement-deconfinement(CD) phase transition line we obtained varies very flatly with chemical potential, which is due to the color potential in Eq.(4) obtained from a pure gluon background that is independent of the chemical potential of quarks. In other words, if the backreaction of quarks on the gluons can be taken into account, the results of CD phase transition may be closer to reality.

\begin{figure}[H]
	\centering
	\includegraphics[width=0.35\textwidth]{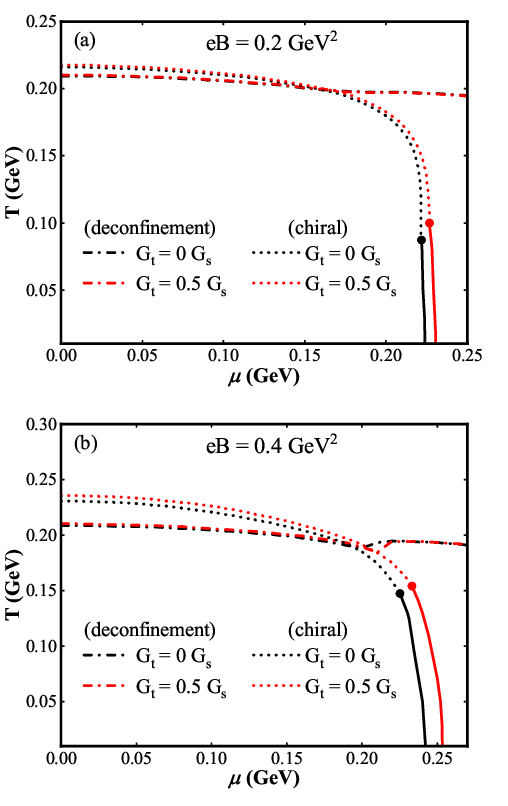}
	\caption{\label{fig4}  $T$-$\mu$  phase diagram for chiral and deconfinement phase transition at different magnetic fields (a) $eB$ = 0.20  and (b)$eB$ = 0.40 GeV$^{2}$ for different spin polarization coupling constants $G_t$ = 0 $G_s$ and $G_t$ = 0.5 $G_s$. The solid lines correspond to chiral first-order phase transition, the dash-dotted lines correspond to chiral crossover phase transition, the full dots correspond to CEP, and the dotted lines correspond to deconfinement crossover phase transition.}
\end{figure}

In addition, according to the specific location of the critical end point (CEP) in Fig. 4, we list the location of the CEP in different cases in Table II. Compared with the case without TSP, we find that the introduction of TSP makes the position of the CEP move slightly. The temperature and chemical potential of the CEP are increased.

\begin{table}[h]
	\caption{CEP position under different magnetic fields and spin polarization coupling constant $G_t$.}
	\begingroup
	\renewcommand*{\arraystretch}{1.5}
	\setlength{\tabcolsep}{8pt} 
	\begin{tabular} {ccc}
		\hline \hline
		&  \( eB = 0.20 \, GeV^2 \)   &  \( eB = 0.40 \, GeV^2  \)  \\
		\hline
	\( G_t = 0\, G_s \) & \( \{T_E, \mu_E\} = \{0.087, 0.222\} \) & \( \{T_E, \mu_E\} = \{0.148, 0.225\} \) \\
	\( G_t = 0.5\, G_s \) & \( \{T_E, \mu_E\} = \{0.102, 0.227\} \) & \( \{T_E, \mu_E\} = \{0.154, 0.233\} \) \\
	\hline\hline
\end{tabular}
\endgroup
\label{Table_parameters}
\end{table}

\subsection{Anisotropic pressure}

The phase diagrams of chiral phase transition and deconfinement phase transition using the PNJL model are shown in Fig. 4. According to this phase diagram, the phase space can be divided into three ranges: (1) the confinement phase with chiral symmetry broken, (2) the deconfinement phase with chiral symmetry restoration, (3) The confinement phase with chiral symmetry restoration (also known as the quarkyonic phase). Next, we study the dependences of the normalized pressures $P_{\parallel}$ and $P_{\bot}$ on the magnetic field after the introduction of TSP in these three ranges, where $P_{\parallel}$ is the pressure parallel to the magnetic field direction, and $P_{\bot}$ is the pressure perpendicular to the magnetic field direction.

\subsubsection{Anisotropic pressure in the confinement phase with chiral symmetry broken}

By considering the influence of TSP, we study the dependence of the transverse pressure and longitudinal pressure on the magnetic field in the confinement phase with chiral symmetry broken in Fig. 5. Fig. 5(a) corresponds to $T$ =0.15 GeV, $\mu$ = 0 GeV. The results show that at small magnetic field $eB$ = 0.01 GeV$^2$, the transverse pressure and longitudinal pressure coincide. When $G_t=0$, the spin polarization effect is not considered, $P_{\parallel}$ and $P_{\bot}$ almost do not change with the increase of the magnetic field, and $P_{\parallel}$ and $P_{\bot}$ are almost the same without splitting. When considering TSP ($G_t=0.5\ G_s$ and $G_t=G_s$),  the longitudinal pressure $P_{\parallel}$ increases not only with the increase of magnetic field, but also with the increase of $G_t$. While the transverse pressure $P_{\bot}$ decreases not only with the increase of magnetic field, but also with the increase of $G_t$. That is to say, the splitting of $P_{\parallel}$ and $P_{\bot}$ not only increases with the increase of magnetic field, but also increases with the increase of $G_t$.

Fig. 5(b) corresponds to $T$ =0.15 GeV, $\mu$ = 0.10 GeV, which is generally similar to that of Fig. 5(a). One finds that when $eB$ =0.01 GeV$^2$, the values of $P_{\parallel}$ and $P_{\bot}$ with $\mu$ =0.10 GeV are larger than that of $\mu$ = 0 GeV, which is the natural result of more quark degrees of freedom \cite{RN56}. On the other hands, it is surprising to find that, when considering TSP, the transverse pressure $P_{\bot}$ produces great oscillation with the increase of magnetic field.

Note that in the confinement phase with chiral symmetry broken, the influence of TSP on the longitudinal and transverse pressure is very significant. In conclusion, considering the contribution of TSP increases the anisotropic characteristics of pressure in the magnetic field background.

The anisotropy of pressure originates from the polarization of quark matter. In our results, increasing the magnetic field strength and the strength of spin-spin interaction will enhance the anisotropy of pressure. This is because, on the one hand, the magnetic field naturally leads to the polarization of quark matter along the direction of the magnetic field. On the other hand, when $G_t\ne 0$, a new tensor spin polarization condensate appears, which is polarized along the direction of the magnetic field. The magnitude of $G_t$ represents the strength of spin-spin interaction, and as $G_t$ increases, TSP increases. In addition, as shown in Fig. 2, increasing the magnetic field also leads to an increase in TSP, resulting in stronger anisotropy. In general, both the magnetic field and $G_t$ enhance the anisotropy of pressure by enhancing the polarization of quark matter. This means that increasing $G_t$ is nearly equivalent to that of the magnetic field.

From Figs. 5(a) and 5(b), we can notice that, as the magnetic field increases, $P_{\bot}$ decreases continuously with the magnetic field and becomes negative at large magnetic fields. This is because the transverse pressure for the interacting case is given in Eq. (14) as $P_{\bot}(T, \mu , eB)=P_{\parallel }(T, \mu , eB)-\mathcal{M}eB$, where there is a competition \cite{ad1} between $P_{\parallel }$ and $\mathcal{M}eB$ . Since the magnetization $\mathcal{M}$ increases steadily with the magnetic field, the transverse pressure tends to decrease and goes to negative values for $T$ = 0.15 GeV at a large value of the magnetic field. Comparing Fig. 5 with Figs. 6 and 7, we find the negative pressure occurs only at low temperature and large magnetic fields, which indicates a paramagnetic squeezing effect \cite{ad2}.  It will lead to the system shrinking in the transverse direction and the density increaseing in the radial direction.

\begin{figure}[H]
	\centering
	\includegraphics[width=0.85\textwidth]{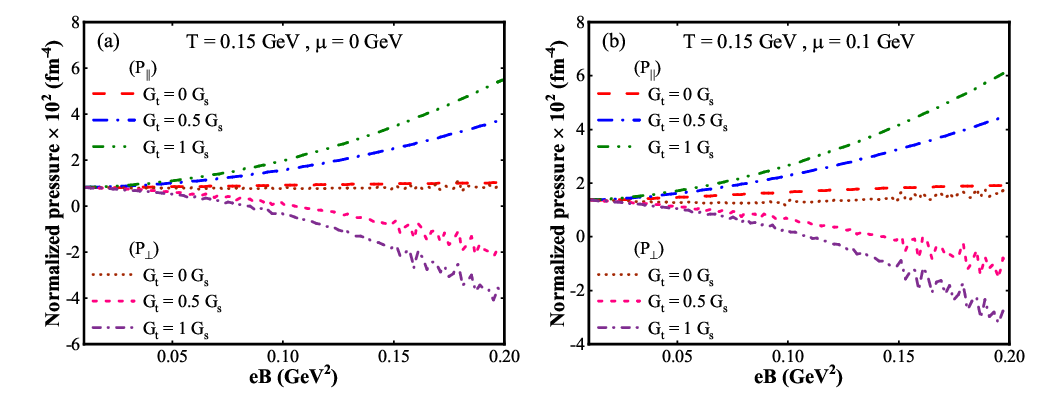}
	\caption{\label{fig5} Normalized longitudinal ($P_{\parallel}$) and transverse ($P_{\bot}$ ) pressure as a function of $eB$ or the confinement phase with chiral symmetry broken for different spin polarization coupling constant $G_t$ with different chemical potential (a) $\mu$ = 0 GeV  and (b) $\mu$ = 0.2 GeV (b), respectively.}
\end{figure}

\subsubsection{Anisotropic pressure in the deconfinement phase with chiral symmetry restored}

By considering the influence of TSP, we study the dependence of the transverse pressure $P_{\bot}$ and longitudinal pressure $P_{\parallel}$ on the magnetic field in the deconfinement phase with chiral symmetry restored shown in Fig. 6. It is found that the longitudinal pressure $P_{\parallel}$ increases with the increase of the magnetic field, and transverse pressures $P_{\bot}$ decrease with magnetic field. An obvious difference from the confinement phase with chiral symmetry broken is that the effect of TSP on anisotropic pressure is very small, and the effect of TSP can be almost ignored. With the restoration of chiral symmetry at high temperature $T$ = 0.25 GeV, the effect of TSP tends to zero.

\begin{figure}[H]
	\centering
	\includegraphics[width=0.85\textwidth]{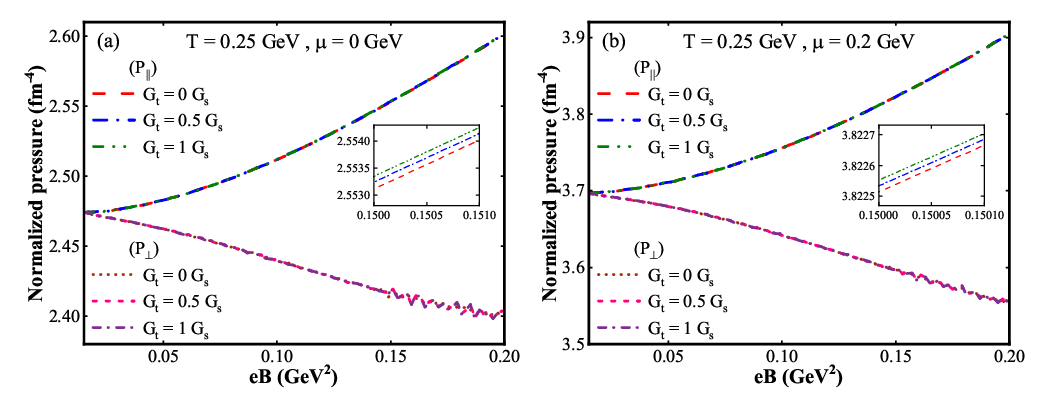}
	\caption{\label{fig6} Normalized longitudinal ($P_{\parallel}$) and transverse ($P_{\bot}$ ) pressure as a function of $eB$ at deconfinement phase with restored chiral symmetry for different chemical potential (a) $\mu$ = 0 GeV  and (b) $\mu$ = 0.2 GeV (b), respectively.}
\end{figure}

\subsubsection{Anisotropic pressure in the quarkyonic phase}

Figure 7 shows the dependences of longitudinal and transverse pressure on magnetic field in quarkyonic phase. The quarkyonic phase is a new phase of QCD, in which the chiral symmetry has been restored, but it is still in the confinement phase \cite{RN33, RN57, RN58, RN59, RN60, RN61, RN62,RN63}. When $T$ =0.15 and $\mu$ = 0.35 GeV, the dynamic quark mass is close to the current quark mass and $\Phi<0.2$, which corresponds to the confinement phase with restored chiral symmetry, meaning that the system is at quarkyonic phase. In this region, the effect of spin polarization condensate on the longitudinal and transverse pressures is very slight, similar to the deconfinement phase with restored chiral symmetry.

The results show that TSP has only a very small effect on the anisotropic pressure in the deconfined phase and the quarkyonic phase with chiral symmetry restored, but it has a very strong effect on the anisotropic pressure in the confined phase with chiral symmetry broken. This is because of the different performance of TSP under different chiral symmetry as we discussed in Fig. 3. The restoration of chiral symmetry means the dissociation of spin polarization condensate. Therefore, attention should be paid to the role of TSP in chiral symmetry breaking. TSP is closely related to the polarization caused by the magnetic field, and the pressure is also affected by the magnetic field. The pressures are slitted into the direction along the magnetic field and perpendicular to the magnetic field, which is a sign of anisotropy. Therefore, when considering the anisotropy of pressure in the chiral symmetry breaking phase, TSP will promote anisotropy.

\begin{figure}[H]
	\centering
	\includegraphics[width=0.5\textwidth]{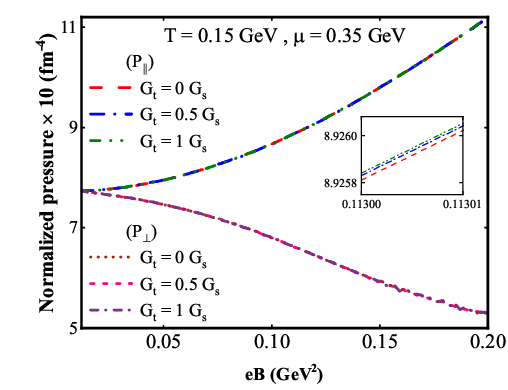}
	\caption{\label{fig7} Normalized longitudinal ($P_{\parallel}$) and transverse ($P_{\bot}$ ) pressure as a function of $eB$ at quarkyonic phase.}
\end{figure}

\section{SUMMARY AND CONCLUSIONS}\label{sec:04 summary}

The generation mechanism of TSP is that quark and antiquark are polarized in the magnetic field background due to the opposite charge and spin, inducing  a magnetic moment along the direction of the magnetic field. This new condensation formed by spin polarization interaction is called the spin polarization condensate.

The introduction of TSP leads to an increase in the dynamical quark mass at the lowest Landau level in the magnetic field background, leading to an increase in the phase transition temperature.  That is to say, for chiral phase transitions, the degree of polarization of quark-antiquark pairs increases with the magnetic field, resulting in an increase in the chiral phase transition temperature. The TSP will correct the position of the CEP. For the deconfinement phase transition, the effect of TSP on the phase transition temperature is very small, and it almost disappears at a large chemical potential.

Because the quark-antiquark pair produces magnetic moments along the direction of the magnetic field, the polarization of the quark-antiquark pair increases with the increase of the magnetic field, resulting in a larger TSP. In the high-temperature QGP background, the pairing of quark and antiquark is blocked, and TSP decreases with the increase of temperature and chemical potential. In addition, since the generation of TSP and chiral condensation depends on the pairing of quark and antiquark, the distribution of TSP is closely related to the chiral phase diagram. Under the same background, the phase transition temperature of TSP is basically the same as the chiral phase transition temperature.

Due to the disappearance of TSP in the chiral symmetry restored phase, but having a large value when the chiral symmetry is broken, we should pay attention to the significant influence of TSP on the pressure anisotropy in the confined phase where the chiral symmetry is broken. TSP can significantly increase the degree of pressure anisotropy in the confined phase.

In this article, we use the PNJL model to discuss the effects of TSP on chiral phase transitions and CD phase transitions with strong magnetic fields. We note that some models \cite{ad3,ad4,ad5} discuss the role of chiral chemical potential $\mu_{5}$  on phase structures, which is an important research direction. We expect to further discuss the characteristics of chiral chemical potential, chiral charge, current, and its susceptibility related to chiral magnetic effects in the future.

\section*{ACKNOWLEDGMENTS}
This work was supported by the National Natural Science Foundation of China (Grants No. 11875178, No. 11475068, No. 11747115).

\section*{References}

\nocite{*}
\bibliography{ref}

\end{document}